\documentclass{article}
\usepackage{spconf,amsmath,graphicx}
\usepackage{amsthm,amsmath,amssymb}
\usepackage{mathrsfs}
\usepackage{booktabs}
\usepackage{threeparttable}
\usepackage{multirow,multicol}
\usepackage{makecell}
\usepackage{cite}

\usepackage[noend]{algpseudocode}
\usepackage{algorithmicx,algorithm}

\title{Improving non-autoregressive end-to-end speech recognition with pre-trained acoustic and language models}
\name{\hspace{-1.7mm}Keqi Deng$^{1,2,*}$, Zehui Yang$^{1,2,*}$, Shinji Watanabe$^{3}$, Yosuke Higuchi$^{4}$, Gaofeng Cheng$^{1}$, Pengyuan Zhang$^{1,2}$
\thanks{$^*$ Equal contribution. This work is partially supported by the National Key Research and Development Program of China (No. 2020AAA0108002).}}
\address{$^1$Key Laboratory of Speech Acoustics and Content Understanding, Institute of Acoustics, CAS, China\\  $^2$University of Chinese Academy of Sciences, China \\
$^3$Carnegie Mellon University, USA, $^4$Waseda University, Japan}
\begin{document}
\ninept
\maketitle
\begin{abstract}
While Transformers have achieved promising results in end-to-end (E2E) automatic speech recognition (ASR), their autoregressive (AR) structure becomes a bottleneck for speeding up the decoding process.
For real-world deployment, ASR systems are desired to be highly accurate while achieving fast inference.
Non-autoregressive (NAR) models have become a popular alternative due to their fast inference speed, but they still fall behind AR systems in recognition accuracy.
To fulfill the two demands, 
in this paper, we propose a NAR CTC/attention model utilizing both pre-trained acoustic and language models: wav2vec2.0 and BERT.
To bridge the modality gap between speech and text representations obtained from the pre-trained models, we design a novel modality conversion mechanism, which is more suitable for logographic languages.
During inference, we employ a CTC branch to generate a target length, which enables the BERT to predict tokens in parallel.
We also design a cache-based CTC/attention joint decoding method to improve the recognition accuracy while keeping the decoding speed fast. 
Experimental results show that the proposed NAR model greatly outperforms our strong wav2vec2.0 CTC baseline ($15.1\%$ relative CER reduction on AISHELL-1).
The proposed NAR model significantly surpasses previous NAR systems on the AISHELL-1 benchmark and shows a potential for English tasks.
\end{abstract}
\section{Introduction}
\label{sec:intro}
End-to-end (E2E) automatic speech recognition (ASR) models simplify the conventional pipeline ASR methods and directly 
convert input speech into corresponding text \cite{8068205,6638947}.
As a way to realize E2E ASR, recurrent neural network transducer (RNN-T) \cite{6638947} and
attention-based encoder-decoder (AED)-based models have been actively studied
\cite{8068205}.
These models are categorized as autoregressive (AR),
which predict a sequence based on a left-to-right chain rule \cite{9414198}. 
On the other hand, non-autoregressive (NAR) models have become popular \cite{9414198,9414694,tian20c_interspeech,Chen_2021} due to its fast inference speed, which can predict tokens simultaneously \cite{9414694,tian20c_interspeech} or iteratively \cite{9414198,Chen_2021}.

In general, AR models cannot be efficiently parallelized during inference,
since their next token generation process depends on previously predicted tokens and
requires incremental computations of a decoder \cite{9414694,Chen_2021}.
Although NAR models can greatly improve the efficiency of inference, they still face two challenges \cite{tian2021tsnat}. The first is to improve recognition performance
since the NAR mechanism often prevents the model from learning the conditional dependencies between output tokens, thus making it fall behind the AR models in performance \cite{Chen_2021,tian2021tsnat}. The second is to improve the training efficiency \cite{9414198,tian20c_interspeech}, as it is difficult to train NAR models and the training convergence is very slow \cite{tian20c_interspeech}.

Among various NAR methods, connectionist temporal classification (CTC) is a popular technique for training a NAR model \cite{10.5555/3044805.3045089}.
CTC achieves a monotonic input-output alignment based on the conditional independence assumption between output tokens.
Recently, a self-supervised training method, wav2vec2.0 \cite{NEURIPS2020_92d1e1eb}, has achieved promising results on CTC models, and the pre-trained model is shown to accelerate the convergence during the fine-tuning stage.
However, even with the pre-trained model obtained by wav2vec2.0, the CTC model needs an external language model (LM) to relax its conditional independence assumption \cite{NEURIPS2020_92d1e1eb,deng21b_interspeech}.
Several works have investigated incorporating BERT into a NAR ASR model to achieve better recognition accuracies \cite{DBLP:journals/corr/abs-2104-04805,9398531,DBLP:conf/icassp/HuangWLCWT21}.
In order to bridge the length gap between the frame-level speech input and token-level text output,
\cite{DBLP:journals/corr/abs-2104-04805} and \cite{9398531}
have introduced global attention and a serial continuous integrate-and-fire (CIF) \cite{9054250}, respectively.
However,
the global attention suffers from poor text length prediction \cite{9398531,dong2020comparison},
and the serial computation in CIF greatly degrades its training efficiency.
Another mismatch lies between the acoustic embedding and the linguistic token embedding of BERT. To solve this,
\cite{DBLP:journals/corr/abs-2104-04805} designs a two-stage training strategy and \cite{9398531} proposes a modal fusion strategy, which both require significantly more training iterations. \par
\begin{figure*}[t]
    \centering
    \includegraphics[width=0.86\linewidth]{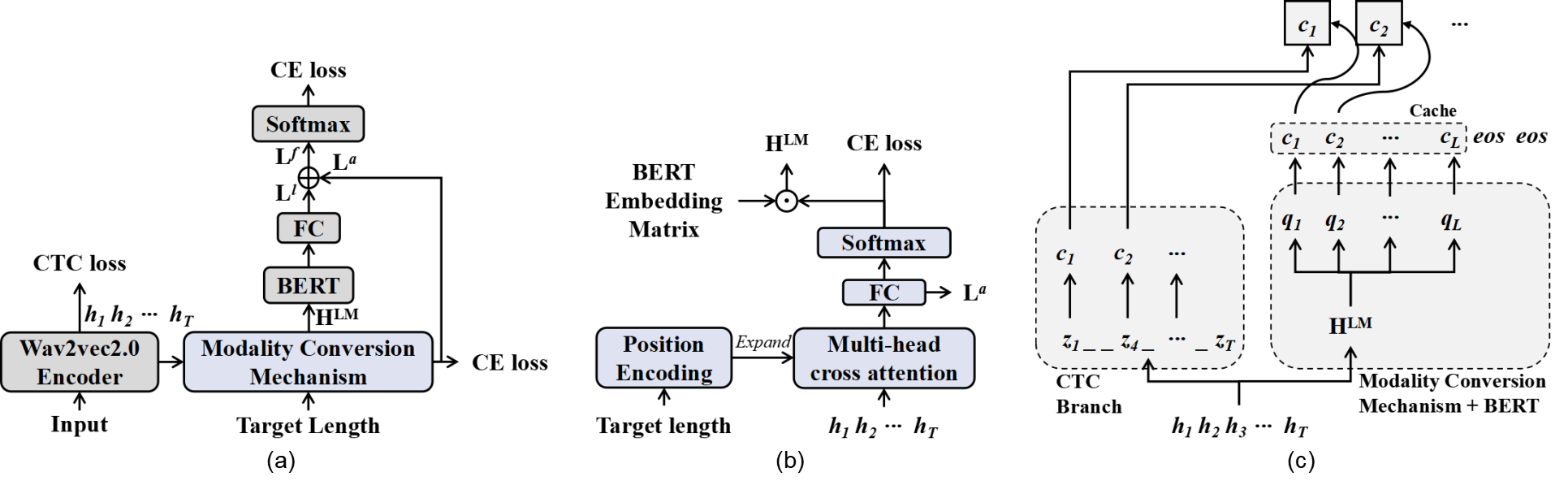}
    \caption{Illustration of our proposed methods. (a) represents the proposed NAR CTC/attention architecture consisting of a wav2vec2.0 encoder and BERT, (b) explains the modality conversion mechanism (MCM), and (c) shows the cache-based CTC/attention joint decoding method.}
    \label{fig:arch}
\end{figure*}
In an attempt to improve training efficiency and recognition performance for NAR models, this paper proposes a novel NAR CTC/attention architecture to fully utilize 
both pre-trained acoustic and language models: wav2vec2.0 and BERT.
In addition, a novel modality conversion mechanism (MCM) is proposed to efficiently bridge the gap between speech and text modalities\footnote{The MCM works well for character-based systems, but needs some improvements for byte pair encoding (BPE)-based \cite{gage1994} systems.}.
Unlike previous methods \cite{DBLP:journals/corr/abs-2104-04805,9398531}, our proposed MCM does not need to greatly increase training iterations. 
To mitigate the length gap between speech and text sequences,
during training, we provide MCM with the ground truth of target length to achieve efficient training. During inference, we employ a CTC branch to generate a target length via greedy search, which enables the BERT to predict tokens in parallel.
To further improve the recognition accuracy while keeping fast decoding speed,
we design a cache-based CTC/attention joint decoding method.
Experimental results show that the proposed model greatly outperforms a strong wav2vec2.0 CTC baseline,
and the cache-based CTC/attention joint decoding method is significantly faster than conventional AR-based beam search.
The proposed model substantially improves over previous NAR systems on the AISHELL-1 benchmark, while its improvement on the Switchboard is not as significant as on the AISHELL-1.

\section{Proposed Methods}
\label{sec:method}
We propose a novel NAR CTC/attention ASR architecture that employs CTC/attention multi-task learning during training.
The realization of our model is shown in Fig.~\ref{fig:arch}, where FC represents a fully connected layer, and \textcircled{·} and \textcircled{+} denote the dot product and addition operations, respectively.
Our model contains a pre-trained wav2vec2.0 encoder as well as a pre-trained BERT and includes a MCM.


\subsection{Wav2vec2.0 Encoder}
\label{ssec:am}
In our proposed NAR CTC/attention model, the CTC branch plays an important role. It is expected to accurately predict a
target length for the modality conversion mechanism and consider all the time boundaries during cache-based CTC/attention joint decoding.
Given the promising results recently achieved by wav2vec2.0 \cite{NEURIPS2020_92d1e1eb, deng2021improving} based on the CTC criterion \cite{CTC_Graves_2006}, we select it as our acoustic encoder.

The wav2vec2.0 encoder consists of a convolutional neural network (CNN)-based feature encoder and a Transformer-based context network. 
As shown in Fig.~\ref{fig:arch} (a),
we use it to extract an acoustic representation ${\rm\mathbf{H}}^{\rm AC}=({\boldsymbol{h}}_1,{\boldsymbol{h}}_2, \cdot\cdot\cdot,{\boldsymbol{h}}_T)^{\top}\in\mathbb{R}^{T\times d}$ from the raw speech waveform $\boldsymbol{x}$, where $d$ is the representation dimension and
$T$ denotes time steps.
During training, we apply a CTC branch after 
the encoder with the CTC objective $\mathcal{L}_{\rm ctc}$.\par

\subsection{Modality Conversion Mechanism and BERT}
\label{ssec:mcm}
Incorporating BERT into an ASR system requires solving the mismatch between speech and text modalities.
As shown in Fig.~\ref{fig:arch} (b),
to efficiently connect acoustic and linguistic representations,
we design a novel structure named modality conversion mechanism (MCM), which consists of the following two parts:\par

First, to bridge the length gap between the acoustic and linguistic representations, we take advantage of the monotonic property of the CTC branch and use it with a greedy search to predict a target length $\hat{L}$.
To achieve efficient training, we choose a ground truth $L$ as our target length during training.
We implement position encoding over this target length $L$ and broadcast its dimension
to be consistent with the acoustic representation ${\rm\mathbf{H}}^{\rm AC}$.
This broadcast is then fed into a cross attention block together with the ${\rm\mathbf{H}}^{\rm AC}$ to capture an alignment between the acoustic representation and its corresponding text. 
This alignment converts the $T$-length ${\rm\mathbf{H}}^{\rm AC}$ to an $L$-length embedding:
\begin{equation}
{\rm \mathbf{H}} = {\rm Attention}({\rm \mathbf{H}}^{\rm PE},{\rm \mathbf{H}}^{\rm AC},{\rm \mathbf{H}}^{\rm AC}),
\end{equation}
where ${\rm \mathbf{H}}^{\rm PE}\in\mathbb{R}^{L\times d}$ denotes the positional embedding, 
and ${\rm \mathbf{H}}\in\mathbb{R}^{L\times d}$ refers to the obtained acoustic embedding.

Second, we attempt to solve the mismatch between the acoustic embedding ${\rm \mathbf{H}}$ and the linguistic token embedding of BERT.
BERT uses very large character/BPE tokens compared with the ASR system, and its adjustment is difficult.
Previous works \cite{DBLP:journals/corr/abs-2104-04805, 9398531}
employ the token embedding of BERT as the learning target of ${\rm \mathbf{H}}$, but these approaches require a significant increase in the training iterations. Further, BERT's token embedding is available during training and decoding, making it unnecessary to
regard it as a learning target.
Therefore, instead of adopting these methods \cite{DBLP:journals/corr/abs-2104-04805, 9398531},
we directly perform a dot product operation on the token embedding of BERT.
Before using BERT, we first generate a preliminary prediction based on the obtained acoustic embedding:
\begin{equation}
{\rm \mathbf{L}}^a = {\rm \mathbf{H}}{\rm \mathbf{W}},
\end{equation}
where matrix ${\rm \mathbf{W}}\in\mathbb{R}^{d\times V}$ are trainable and ${\rm \mathbf{L}}^a\in\mathbb{R}^{L\times V}$ denotes the logits of acoustic embedding. Letting
$V$ denote the vocabulary size of the ASR system, we then have
${\rm \mathbf{M}_{\rm BERT}}\in\mathbb{R}^{V\times d}$, a matrix that contains the token embedding of BERT shared with the ASR system,
arranged in the order of
tokens list in the ASR system\footnote{It should be noted that this approach works well with character-based BERT for Mandarin, but it needs extensions for BPE tokens used in English as the ASR system and pre-trained LM system have different BPE models, which are related to the vocabulary size of BPE thus cannot be shared.}.
We can then achieve an efficient modality conversion through dot product: 
\begin{eqnarray}
{{\rm \mathbf{W}}^a}&=&{\rm Softmax}({\rm \mathbf{L}}^a),\\
{\rm \mathbf{H}}^{\rm LM}&=&{{\rm \mathbf{W}}^a}\cdot{\rm \mathbf{M}_{\rm BERT}},
\end{eqnarray}
where ${\rm \mathbf{W}}^a\in\mathbb{R}^{L\times V}$ and each element ${\rm \mathbf{W}}^a_{ij}$ denotes the weight assigned to the $j$-th BERT token embedding at the $i$-th step.
We then obtain
the linguistic representation ${\rm \mathbf{H}}^{\rm LM}\in\mathbb{R}^{L\times d}$, which is fed into the BERT encoder. 
We use a cross-entropy (CE) criterion $\mathcal{L}_{ce1}$ to encourage the ${\rm \mathbf{L}}^a$ after softmax to generate correct predictions before feeding it into the BERT.


We choose to incorporate BERT \cite{DevlinCLT19} into our ASR system due to its
powerful text processing capabilities enabled by its 
embedding layer 
and a multi-layer Transformer encoder 
\cite{DevlinCLT19, 9664007}. 
As shown in Fig.~\ref{fig:arch} (a), we apply a FC layer on the top of the BERT output to obtain logits of the linguistic representation ${\rm \mathbf{L}}^l$.


\subsection{Training Objective}
\label{ssec:train}

To utilize both acoustic and linguistic knowledge, we add the ${\rm \mathbf{L}}^a$ and ${\rm \mathbf{L}}^l$ together to get the final logits
${\rm \mathbf{L}}^f$:
\begin{equation}
{\rm \mathbf{L}}^f = \alpha{\rm \mathbf{L}}^l+{\rm \mathbf{L}}^a, \label{logits}
\end{equation}
where $\alpha$ is a tunable hyper-parameter. We calculate the CE loss $\mathcal{L}_{\rm ce}$ over the final output ${\rm \mathbf{L}}^f$ to provide further supervision.\par

During joint training, the loss function is defined by:
\begin{equation}
	\mathcal{L} = (1-\beta)(\lambda_1\mathcal{L}_{\rm ce} +\lambda_2\mathcal{L}_{\rm ce1})+\beta\mathcal{L}_{\rm ctc}, 
	\label{loss}
\end{equation}
where $\lambda_1$, $\lambda_2$, and $\beta$ are tunable hyper-parameters. $\mathcal{L}_{\rm ctc}$ and $\mathcal{L}_{\rm ce1}$ are defined in Sections~\ref{ssec:am} and~\ref{ssec:mcm}, respectively.
\begin{algorithm}[t]
\caption{Cache-based joint CTC/attention decoding} 
\footnotesize
\begin{algorithmic}[1]
\State $\Omega_0\gets \emptyset$ 
\State $\hat{\Omega}\gets \emptyset $
\State $\hat{L}\gets$ target length generated by CTC greedy search
\For{$l=1\cdot\cdot\cdot L_{\max}$}
    \State $\Omega_l\gets\emptyset$
    \While {$\Omega_{l-1} \neq \emptyset\ \rm{or}\ l=1 $}
        \State $g\gets \mathrm{HEAD}(\Omega_{l-1})$
        \State DEQUEUE$(\Omega_{l-1})$
        \For {each $c \in \mathcal{V}\cup{\left \langle eos\right \rangle}$}
            \State $h\gets g\cdot c$
            \If{$l>\hat{L}$}
                \State $\alpha_{f}(c=<eos>,x)\gets\log(0.9)$
                \State $\alpha_{f}(c\neq<eos>,x)\gets\log(0.1/{V})$
            \EndIf
            \State $\alpha(h,x)\gets\mu\alpha_{ctc}(h,x)+(1-\mu)\alpha_{f}(c,x)$
            \If{$c=<eos>$}
                \State ENQUEUE$(\hat{\Omega},h)$
            \Else
                \State ENQUEUE$(\Omega_l,h)$
                \If{$|\Omega_l|>beamWidth$}
                    \State REMOVEWORST$(\Omega_l)$
                \EndIf
            \EndIf
        \EndFor
    \EndWhile
    \If{ENDDETECT$(\hat{\Omega},l)=\mathrm{true}$}
        \State \rm break
    \EndIf
\EndFor
\State \Return $\arg\max_{h\in\hat{\Omega}}\alpha(h,x)$
\end{algorithmic}
\end{algorithm}
\subsection{Cache-based CTC/attention Joint Decoding}
\label{ssec:dec}
We employ the CTC branch to generate a target length $\hat{L}$, 
which enables the BERT to predict tokens in parallel. 
Inspired by \cite{hori-etal-2017-joint}, 
we design a cache-based CTC/attention joint decoding method to further improve the recognition accuracies, as shown in Fig.~\ref{fig:arch} (c). Unlike the conventional CTC/attention joint decoding \cite{hori-etal-2017-joint}, we precompute the attention-based score up to $\hat{L}$ length at one step instead of incremental computation, thus enabling our ASR model to achieve high inference speed
\footnote{Strictly speaking, this is not a NAR decoding, but it is used to further improve the proposed model with efficient computation.}.
\par 

        


The details of our method are shown in
Algorithm 1, where $\Omega_l$ 
and $\hat{\Omega}$ denote queues that accept
partial hypotheses of length $l$ and complete ones, respectively. 
$\mathcal{V}$ is the token vocabulary with $V$ size. Each $c \in \mathcal{V}\cup{\left \langle eos\right \rangle}$ is appended to a former partial hypothesis $g$ given by $\Omega_{l-1}$. 
The attention-based scores of $\hat{L}$ length sequence are calculated at one step and then stored in the cache. When $l>\hat{L}$, we predict ${\left \langle eos\right \rangle}$ with a 0.9 probability. 
We score each extended hypothesis $h\mbox{=}g\cdot c$  by a CTC prefix score and attention-based scores stored in the cache, as shown in line 14. This hypothesis is stored in either $\Omega_l$ or $\hat{\Omega}$ based on the value of $c$. While $c\neq{\left \langle eos\right \rangle}$, $h$ is added to $\Omega_l$, and $\Omega_l$ 's size is then compared with the beam width for pruning by REMOVEWORST($\cdot$). We also utilize an additional function ENDDETECT$(\hat{\Omega},l)$ \cite{hori-etal-2017-joint} to determine whether to stop the procedure before $l$ reaches $L_{\max}$. If finding $h\in\hat{\Omega}$ with higher scores is almost impossible as $l$ increases, ENDDETECT$(\hat{\Omega},l)$ returns $true$. \par

We apply softmax to ${\rm \mathbf{L}}^f$ to get an attention-based probability $p_f(c|x)$. 
We also cumulatively calculate a prefix CTC probability $p_{\rm ctc}(h,\cdot\cdot\cdot|x)$. The scores
are then computed in log domain:
\begin{equation}
    \left \{
    \begin{aligned}
        &\alpha_{\rm ctc}(h,x) = \log p_{\rm ctc}(h,\cdot\cdot\cdot|x)\\
        &\alpha_f(c,x) = \log p_f(c|x)
    \end{aligned}
    \right.
\end{equation}


\section{Experiments}
\label{sec:exp}

\subsection{Corpus}
\label{ssec:corp}
We evaluate our proposed NAR CTC/attention model on the Mandarin AISHELL-1 \cite{8384449} and English Switchboard \cite{225858} corpora.
We also use an unlabeled speech training set from the AISHELL-2 corpus to pre-train a Mandarin wav2vec2.0 base model \cite{DBLP:journals/corr/abs-1808-10583}.


\subsection{Model Descriptions}
\label{ssec:model}
The ESPnet2 toolkit \cite{watanabe2018espnet} is used to build our wav2vec2.0 CTC baseline and our proposed NAR CTC/attention model. 
We employ raw speech as the acoustic input. 
For Mandarin, we use 4230 Chinese characters with 3 non-verbal symbols: blank, unknown-character, and sos/eos
as the modeling units. For English, we utilize 2000 modeling units, including 1997 BPE \cite{gage1994} units and 3 non-verbal symbols.

Our wav2vec2.0 CTC baseline consists of a base wav2vec2.0 \cite{NEURIPS2020_92d1e1eb} encoder 
and a FC layer as the classifier with a size of 4233 and 2000 for Mandarin and English tasks respectively. 
If not specified, wav2vec2.0 refers to
Mandarin wav2vec2.0 in the Mandarin task.
As for our proposed NAR CTC/attention model, the wav2vec2.0 encoder and CTC branch are the same as those in the baseline model, and its MCM contains one-layer multi-head cross attention with 768 model dimensions and 4 heads. The FCs in Fig.~\ref{fig:arch}(a) and Fig.~\ref{fig:arch}(b) are of the same size as the CTC branch.
The pre-trained Mandarin BERT (i.e., bert-base-chinese) and the English RoBERTa
\footnote{RoBERTa is a variant of BERT, and we choose it for our English tasks because its modeling unit is BPE and thus more suitable for our English tasks.}
\cite{DBLP:journals/corr/abs-1907-11692} (i.e., roberta-base)
are both provided by Huggingface Transformer Library \cite{wolf-etal-2020-transformers}, and the English wav2vec2.0 base model is provided by Fairseq \cite{ott2019fairseq}.
For the first 5000 and 25000 steps, 
the parameters of both the wav2vec2.0 and the BERT/RoBERTa are fixed. 
In our embedding layer and the first 3 and 6 Transformer layers of BERT and RoBERTa, the parameters
are always fixed.
We set both the $\lambda_1$ and $\lambda_2$ in Eq.~\eqref{loss} to 0.5. The $\beta$ in Eq.~\eqref{loss} and $\alpha$ in Eq.~\eqref{logits} are set to 0.3. Real time factor (RTF) is measured on the test set of AISHELL-1 using a P100 GPU. 

Following the ESPnet2 recipe \cite{watanabe2018espnet},
we fine-tune the Mandarin BERT and English RoBERTa with subsequent masks
as external AR-based LMs for the baseline. During inference, we set the beam size as 10 and the weight of BERT/RoBERTa LM is 0.3 when used.

\subsection{Main Results}
\label{ssec:res}
\begin{table}[t]
  \caption{The character error rates (CER) (\%) of different AR/NAR ASR models on AISHELL-1 corpus.}
  \label{tab:ar-nar}
  \centering
  \setlength{\tabcolsep}{1.3mm}
  \begin{threeparttable} 
  \begin{tabular}{c l c c c}
    \toprule
    \multicolumn{2}{c}{Model}  & Epoch & Dev & Test \\
    \midrule
    \multicolumn{2}{l}{\emph{Autoregressive ASR}} & & & \\
    & resGSA-Transformer \cite{liang21_interspeech}&50&5.4&5.9\\
    & ESPnet (Transformer) \cite{watanabe2018espnet} &50& 5.9 & 6.4 \\ 
    & ESPnet (Conformer) \cite{watanabe2018espnet} &50& 4.4 & 4.7 \\
    \cmidrule(lr){2-5}
    & W2v2 CTC baseline + external BERT LM &20& 4.4 & 4.7 \\
    \midrule
    \multicolumn{2}{l}{\emph{Non-autoregressive ASR}} & & & \\
    & NAR-Transformer \cite{9414694} &200&5.3 &5.9 \\
    & A-FMLM \cite{Chen_2021} &50&6.2&6.7\\
    & TSNAT-Big+Two Step Inference \cite{tian2021tsnat}  &100&5.1 &5.6 \\
    &D-Att shared Enc \cite{9362115} &50& -- &6.5\\
    & NAR-BERT-ASR \cite{DBLP:journals/corr/abs-2104-04805} &130&4.9&5.5\\
    &CASS-NAT \cite{fan2021cass}&90&5.3&5.8\\
    &LASO-big with BERT \cite{9437636}&130&5.2&5.8\\
    \cmidrule(lr){2-5}
    & W2v2 CTC baseline &20& 4.8 & 5.3 \\ 
     & Proposed NAR CTC/attention &\textbf{20}& \textbf{4.1} & \textbf{4.5} \\
    \bottomrule
  \end{tabular}
\end{threeparttable}  
\end{table}
We compare the performance of our proposed NAR CTC/attention model with the strong wav2vec2.0 CTC baseline and other AR/NAR systems. The results are shown in Table~\ref{tab:ar-nar}, where W2v2 denotes the Mandarin wav2vec2.0. 
It should be noted that when the CTC baseline model uses an external BERT LM during one-pass decoding with beam search, we classify it as an AR system because the BERT is set to a unidirectional AR structure during 
the fine-tuning stage. 
And our NAR CTC/attention model 
employs greedy search during inference.
The results show that our proposed NAR CTC/attention model 
greatly outperforms the 
strong wav2vec2.0 CTC baseline, yielding $15.1\%$ relative CER reduction.

Furthermore, even when a strong BERT-based AR LM is employed for the
baseline to relax its conditional independence assumption, the results show that our proposed NAR CTC/attention model still achieves better recognition accuracy in addition to maintaining the advantage of the NAR model's fast inference speed, which is not achieved by previous works \cite{9398531,DBLP:journals/corr/abs-2012-12121}.
Comparing with other AR/NAR systems, our NAR model greatly improves over previous NAR systems with much fewer training epochs and it even exceeds the AR-based systems. 
Therefore, it can be concluded that our NAR CTC/attention model achieves impressive performance with high training efficiency.

\subsection{Ablation Studies on Cache-based CTC/attention Decoding}
\label{ssec:ab-1}
In the main experiments, we employ the greedy search for our NAR model.
In this section, we conduct ablation studies to verify the effectiveness of our cache-based CTC/attention joint decoding. The results are shown in Table~\ref{tab:dec}, where Baseline represents our wav2vec2.0 CTC baseline and
Cache CTC/Att denotes the cache-based CTC/attention joint decoding (i.e. beam search).
\begin{table}[t]
  \caption{The CER (\%) of our NAR CTC/attention system with different decoding style on AISHELL-1 corpus.}
  \label{tab:dec}
  \centering
  \setlength{\tabcolsep}{1.3mm}
  \begin{tabular}{l c c c c c}
    \toprule
    \multirow{2}{*}{Model}&
    {Wav2vec2.0} &
    \multirow{2}{*}{{Decoding Style}}&
    \multirow{2}{*}{{Dev}} &
    \multirow{2}{*}{{Test}}&\multirow{2}{*}{{RTF}}\\
     & {Encoder} & & & \\
    \midrule
    Baseline & English & Greedy search &6.1 & 6.5&0.016\\
    + BERT LM & English & Beam search &5.4 & 5.7&8.098\\
    Baseline & Mandarin & Greedy search&4.8 & 5.3&\textbf{0.015}\\
    + BERT LM & Mandarin & Beam search &4.4 & 4.7&8.098\\
    \midrule
    Proposed & English & Greedy search&5.1 & 5.8&0.044\\
    Proposed & English & Cache CTC/Att  &4.8 & 5.1&0.665\\
    Proposed & Mandarin & Greedy search &4.1 & 4.5&0.045\\
    Proposed & Mandarin & Cache CTC/Att &\textbf{4.0} & \textbf{4.3}&0.665\\
    \bottomrule
  \end{tabular}
\end{table}

It can be seen from the results that the conclusions are the same whether we use the Mandarin or the English wav2vec2.0 encoder:
1) Our proposed NAR model outperforms the baseline even with vanilla attention-based greedy search. 
2) After using the cache-based CTC/attention joint decoding method, our model has been further improved, especially on the English wav2vec2.0 encoder.
3) Although the cache-based CTC/attention joint decoding is slower compared to greedy search, it is significantly faster than the beam search with an AR-based external BERT LM, as our model only needs to be run for one time during inference.

In addition, the improvement on the English wav2vec2.0 encoder indicates that
our proposed method may also help assist ASR tasks in low-resource languages based on cross-lingual pre-training.
\subsection{Experimental Results on Alphabetic Language}
\label{ssec:ab-3}
Previous experiments have verified the effectiveness of our proposed model on Mandarin (i.e. logographic language) tasks.
However, if we consider applying our method to alphabetic languages (e.g., English),
it raises a problem regarding the vocabulary size of BPE,
as an ASR system and BERT are likely to have different output units.
For example, given the sentence "$good\ weather$", the tokenized input for RoBERTa is "$\_good\ \_weather$", but the corresponding input for the ASR system is "$\_good\ \_wea\ ther$". 
The BPE vocabulary size of RoBERTa is often word-level,
which makes it too large and sparse for the ASR system to train on, especially for the CTC branch.
A simple solution is that pre-training RoBERTa with the same BPE size as the ASR system.
However, it is not always feasible due to the requirement of large computational resources. 

To solve this challenge, first, since BPE units like "$\_wea$" and "$ther$" also exist in the token list of RoBERTa, we can directly convert
these units
to their corresponding index in RoBERTa's token list before entering into the RoBERTa model. Second, we can also use CTC greedy search results as input to the MCM during training and decoding to avoid the length mismatch. 


We choose the English Switchboard \cite{225858} corpus to evaluate our NAR model on alphabetic language.
The results in Table~\ref{tab:ar-nar-en} show that our method outperforms the strong wav2vec2.0 CTC baseline and AR-based Transformer system. However, after using a RoBERTa-based AR LM, the baseline surpasses our NAR model.
\renewcommand{\arraystretch}{0.8}
\begin{table}[t]
  \caption{The word error rate (WER) (\%) of different AR/NAR ASR models on Switchboard corpus.}
  \label{tab:ar-nar-en}
  \centering
  \setlength{\tabcolsep}{1.3mm}
  \begin{threeparttable} 
  \begin{tabular}{c l c c c}
    \toprule
    \multicolumn{2}{c}{{Model}}  & {Epoch} & {Dev} & {Eval2000} \\
    \midrule
    \multicolumn{2}{l}{\emph{{Autoregressive ASR}}} & & & \\
    & ESPnet (Transformer) \cite{watanabe2018espnet} &100& -- & 12.9 \\ 
    & ESPnet (Conformer) \cite{watanabe2018espnet} &150& -- & \textbf{10.4} \\
    \cmidrule(lr){2-5}
    & W2v2 CTC baseline + RoBERTa LM &25& \textbf{10.0} & 10.8 \\
    \midrule
    \multicolumn{2}{l}{\emph{{Non-autoregressive ASR}}} & & & \\
    & W2v2 CTC baseline &25& 11.3 & 12.4 \\ 
     & Proposed NAR CTC/attention &25& 10.6 & 11.9 \\
    \bottomrule
  \end{tabular}
\end{threeparttable}  
\end{table}

\section{Conclusion}
\label{sec:con}
In this paper, 
we proposed a NAR CTC/attention model that fully utilizes the pre-trained wav2vec2.0 and BERT.
We also designed a novel modality conversion mechanism (MCM) to efficiently bridge the gap between speech and text modalities. During inference, we made use of a CTC branch to generate a target length, which enables BERT to process tokens in parallel. Furthermore, we proposed a cache-based CTC/attention joint decoding method to further improve the recognition accuracy while keeping the decoding speed fast.
The experimental results showed that the proposed model improves over our wav2vec2.0 CTC baseline and other NAR models.
\pagebreak

\bibliographystyle{IEEEbib}
\bibliography{strings,refs}

\end{document}